\documentclass[twocolumn,showpacs,twoside,10pt,prl,superscriptaddress]{revtex4-1}

\usepackage{graphicx}
\usepackage{dcolumn}
\usepackage{bm}

\usepackage{amsmath,amsfonts}
\usepackage{algorithmic}
\usepackage{algorithm}
\usepackage{array}
\usepackage[caption=false,font=normalsize,labelfont=sf,textfont=sf]{subfig}
\usepackage{textcomp}
\usepackage{stfloats}
\usepackage{url}
\usepackage{verbatim}
\usepackage{braket}
\usepackage{amsmath}
\usepackage{graphicx}
\usepackage{subfig}
\usepackage[backref=true]{hyperref}

\emergencystretch=\maxdimen
\begin{document}
	
	\preprint{APS/123-QED}
	
	\title{A Fully-integrated Diamond Nitrogen-Vacancy Magnetometer \\with Nanotesla Sensitivity}
	
	\author{Yulin Dai}
	\affiliation{Institute of Quantum Sensing and College of Optical Science and Engineering, \\ Zhejiang University, Hangzhou, 310027. China}
	\affiliation{Research Center for Quantum Sensing, Zhejiang Lab, Hangzhou, 311000, China}
	
	\author{Wenhui Tian}
	\affiliation{Research Center for Quantum Sensing, Zhejiang Lab, Hangzhou, 311000, China}
	\author{Qing liu}
	\affiliation{Research Center for Quantum Sensing, Zhejiang Lab, Hangzhou, 311000, China}
	
	\author{Bao Chen}
	\affiliation{Institute of Quantum Sensing and College of Optical Science and Engineering, \\ Zhejiang University, Hangzhou, 310027. China}
	\affiliation{Research Center for Quantum Sensing, Zhejiang Lab, Hangzhou, 311000, China}
	
	\author{Yushan Liu}
	\affiliation{Research Center for Quantum Sensing, Zhejiang Lab, Hangzhou, 311000, China}
	
	\author{Qidi Hu}
	\affiliation{Research Center for Quantum Sensing, Zhejiang Lab, Hangzhou, 311000, China}
	
	\author{Zheng Ma}
	\affiliation{Research Center for Quantum Sensing, Zhejiang Lab, Hangzhou, 311000, China}
	
	\author{Yunpeng Zhai}
	\affiliation{Institute of Quantum Sensing and College of Optical Science and Engineering, \\ Zhejiang University, Hangzhou, 310027. China}
	
	\author{Haodong Wang}
	\affiliation{Institute of Quantum Sensing and College of Optical Science and Engineering, \\ Zhejiang University, Hangzhou, 310027. China}
	
	\author{Ying Dong}
	\affiliation{Research Center for Quantum Sensing, Zhejiang Lab, Hangzhou, 311000, China}
	
	\author{Nanyang Xu}
	\email{nyxu.physics@zju.edu.cn}
	\affiliation{Research Center for Quantum Sensing, Zhejiang Lab, Hangzhou, 311000, China}
	\affiliation{Institute of Quantum Sensing and College of Optical Science and Engineering, \\ Zhejiang University, Hangzhou, 310027. China}

	\date{\today}

	\begin{abstract}
		
		Ensemble diamond nitrogen-vacancy (DNV) centers have emerged as a promising platform for precise earth-field vector magnetic sensing, particularly in applications that require high mobility. Nevertheless, integrating all control utilities into a compact form has proven challenging, thus far limiting the sensitivity of mobile DNV magnetometers to the $\mu$T-level. This study introduces a fully integrated DNV magnetometer that encompasses all the essential components typically found in traditional platforms, while maintaining compact dimensions of approximately $\Phi$ 13 cm × 26 cm. In contrast to previous efforts, we successfully address these challenges by integrating a high-power laser, a lock-in amplifier, and a digitally-modulated microwave source. These home-made components show comparable performance with commercial devices under our circumstance, resulting in an optimal sensitivity of 2.14 nT/$\sqrt{Hz}$. The limitations in this system as well as possible future improvements are discussed. This work paves the way for the use of DNV magnetometry in cost-effective, mobile unmanned aerial vehicles, facilitating a wide range of practical applications.
		
	\end{abstract}
	
	\maketitle
	
	\section{Introduction}
	Diamond nitrogen-vacancy (DNV) magnetometers \cite{ref4,ref5,ref6,ref1.1.2} offering sensitivities up to sub-picotesla level \cite{ref5, ref1.1.4, ref1.1.6, ref1.1.8, ref1.1.9, ref1.1.10, ref1.2.1, ref1.2.2, ref1.2.4, ref1.2.5, ref1.2.7, ref1.2.8, ref1.2.9, ref1.2.11, ref1.7.1, ref1.7.3, ref1.7.4, ref1.7.5, ref1.7.6, ref1.7.7, ref1.7.8, ref1.7.9, ref1.7.10,ref1.7.11,ref1.7.12,ref1.7.14,ref1.7.15,ref1.7.16, ref1.7.17}, providing precise vector magnetic sensing without heading errors, have great potential in biomedical detection \cite{ref1.1.2,ref1.1.4,ref1.1.6,ref1.1.8,ref1.1.9,ref1.1.10}, material studying \cite{ref1.2.1,ref1.2.2,ref1.2.4,ref1.2.5,ref1.2.7,ref1.2.8,ref1.2.9,ref1.2.11} and navigation \cite{ref1.3.1}. 
	Advancements in miniaturized DNV magnetometers \cite{ref1.10.1,ref1.10.2,ref1.10.3,ref1.11.1,ref1.11.2,ref1.11.3,ref1.11.4,ref1.11.6,ref1.12.1,ref1.9.1,ref1.9.2,ref1.9.3,ref1.8.1,ref1.8.2} have offered alternatives to classical platforms due to their compact and portable nature, making them suitable for a wide range of applications. These magnetometers can be classified based on functional completeness into portable probes and fully-integrated devices. Up till now, most developments have focused on integrating fiber-based probes \cite{ref1.10.1,ref1.10.2,ref1.10.3,ref1.11.1,ref1.11.2,ref1.11.3,ref1.11.4,ref1.11.6,ref1.12.1} by attaching a diamond at the end of a fiber, to exhibit high sensitivities ranging from 310pT/$\sqrt{Hz}$ to 180nT/$\sqrt{Hz}$, compact sizes from fiber dimensions to several centimeters, and high spatial resolutions suitable for medical imaging. Some works are approaching the chip-size integration \cite{ref1.9.1,ref1.9.2,ref1.9.3} utilizing nano-plasmonic structures on complementary metal-oxide-semiconductor (CMOS) platforms to perform various functions with external lasers. The chips demonstrate compact footprints ranging from 0.04mm$^{2}$ to 1.5mm$^{2}$ and achieve sensitivities ranging from 245nT/$\sqrt{Hz}$ to 32.1uT/$\sqrt{Hz}$. 
	
	The advancements above have contributed to the applications of DNV magnetometers in labs or fixed places. While on mobile platforms, DNV magnetometer with full functionalities \cite{ref1.8.1, ref1.8.2} suffer from heating problems from laser and microwave amplifier, and the absence of on-board electronics for modulated microwave and lock-in detection. These limitations have restricted their sensitivities to the order of $\mu$T/$\sqrt{Hz}$ \cite{ref1.8.2}, far from the requirement of nT-level in navigation and surveying applications \cite{ref1.13.1}. Furthermore, the magnetometer with low Size, Weight, and Power (SWaP) is still required, yet has not been reported based on DNV platforms for mobile cost-effective vector magnetic-field sensing, for example, on the mobile unmanned aerial vehicles.

	Here we present a portable DNV magnetometer encompassing all the functionalities of a conventional platform. To achieve high-sensitivity vector sensing within a compacted volume ($\Phi$13cm×26cm), we have overcome the challenges of individual components and adopt a fully-integrated design of hardware based on our former works, including a measurement controller with a lock-in amplifier (LIA) \cite{ref3.1.4,ref3.1.3}, a DDS-based digital microwave (MW) source \cite{ref3.2.1} a probe with high-power laser and other supporting components. We evaluated the performance of all essential components and found that they are comparable with those of commercial instruments. Combination with previous optimization techniques such as simultaneous multi-frequency microwave and digital balance detection, we achieve a sensitivity ranging from 2.14 to 4.58 nT/$\sqrt{Hz}$ associated with an adjustable detection frequency bandwidth from 1.3 to 625 Hz. The magnetometer presented is well-suited for installation on unmanned aerial vehicles (UAVs), enabling mobile applications such as autonomous navigation and mineral survey.
	
	\begin{widetext}
		\begin{minipage}{0.96\linewidth}
			\begin{figure}[H]
				\centering
				\includegraphics[width=\linewidth]{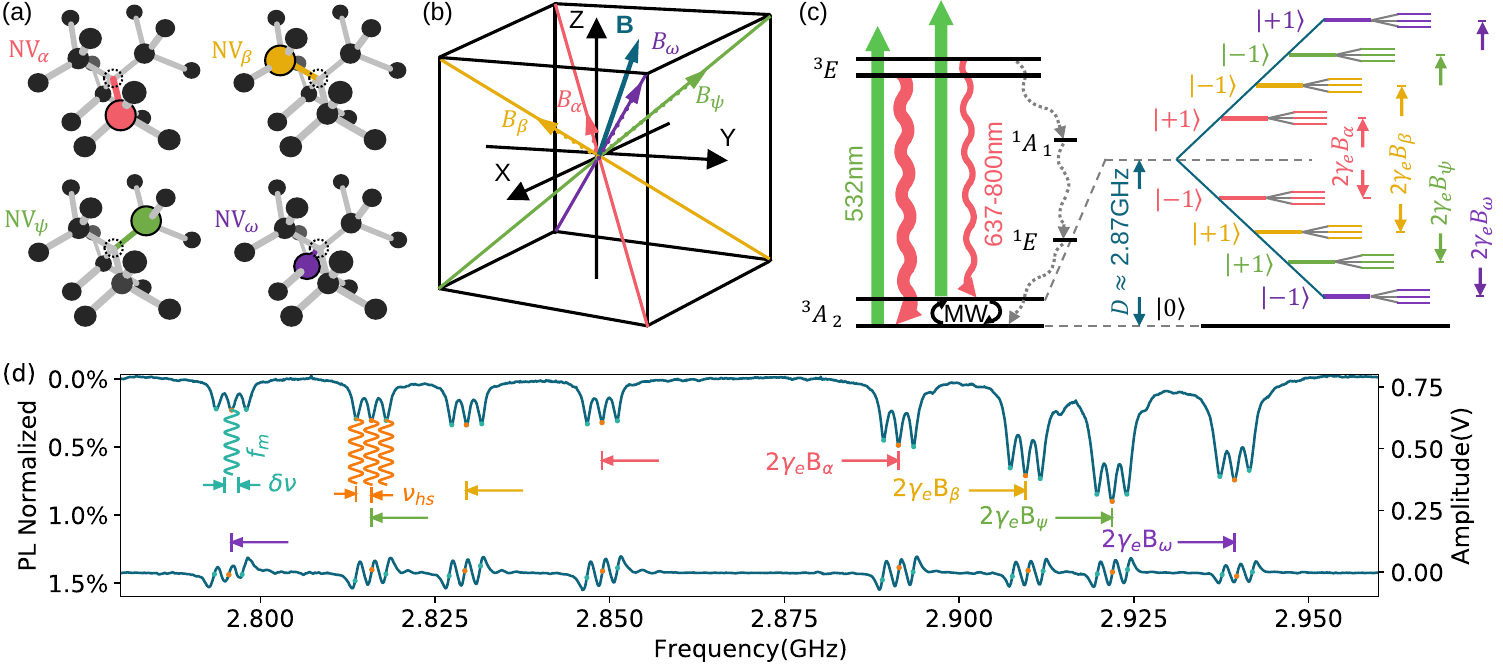}
				\caption{(a) Diamond lattice structure with four distinct NV centers having symmetry axes oriented along different orientations: $NV_{i}$ ($i=\alpha,\beta,\psi,\omega$). (b) The external magnetic field B projected onto the four symmetry axes as $B_{i}$. (c) Energy level diagram of a negatively charged NV$^{-}$ center in diamond. In the presence of an external magnetic field, the $\ket{\pm1}$ sublevel of the electronic spin in various NV centers experiences a Zeeman shift to different energy levels of 2$\gamma_e B_{i}$. (d) Continuous wave ODMR spectrum and lock-in signal. Microwaves used for lock-in technique are shown as sinusoidal waves: a modulated microwave with a modulation frequency $f_m$ and a modulation depth $\delta\nu$, and a multi-frequency microwave mixed with a side-band wave $\nu_{hs}$}
				\label{fig1}
			\end{figure}    
		\end{minipage}
		
	\end{widetext}

	\section{RESULT}
	As shown in Fig.\ref{fig1}a-c, nitrogen-vacancy (NV) center in diamond is a crystal point defect that consists of a substitutional nitrogen atom and an adjacent vacancy in a carbon lattice. The spin energy levels of the NV$^{-}$ defects are highly sensitive to the magnetic fields projected to the four directional NV symmetry axes, which are along four [111] crystallographic orientations. The NV$^{-}$ center exhibits an electronic spin ground triplet state that contains $\ket{m_{s}=0}$ and $\ket{m_{s}=\pm1}$ sublevels with a zero-field splitting D $\approx$ 2.87GHz. Exposed in an external magnetic field B, the $\ket{\pm1}$ sublevel experiences Zeeman shift and splits to eight sublevels, each contains three hyperfine subfeatures that arise from interactions between the NV$^{-}$s and $^{14}$N nuclear spins at 2.158 MHz. The magnetic-dependent transition frequencies is given by $\Delta\nu_{\alpha}=2\gamma_e B_{i}$, where  $\gamma$$_{e}$ is the gyro-magnetic ratio of NV electron spin and $i=\alpha,\beta,\psi,\omega$ denotes the direction of the centers. This makes ensemble NV centers an ideal platform for simultaneously vector magnetic-field measurement.
	
	When excited by a 532nm green laser, the electronic spins of NV center initially transit from the ground state $^{3}$A$_{2}$ to the excited state $^{3}$E. Most of the spins then return to the ground state by emitting red fluorescence within the 637-800nm wavelength range, while some of the $\ket{\pm1}$ state spins undergo non-radiative decay via an inter-system crossing that involves meta-stable singlet states $^{1}$A$_{1}$ and $^{1}$E before returning to the $\ket{0}$ ground state. As results, the photoluminescence (PL) of the $\ket{0}$ state is brighter than that of the $\ket{\pm1}$ state, and after lots of transition cycles, the spin population polarizes into the $\ket{0}$ state. The two spin states can be converted by applying microwaves at the transition frequencies. By sweeping a single microwave tone over a suitable frequency range, the transition frequencies can be detected via the optically detected magnetic resonance (ODMR) spectrum as shown in Fig.\ref{fig1}d (the upper spectrum). If the microwave is modulated with a modulation frequency $f_m$ and a modulation depth $\delta\nu$, lock-in detection can be applied to get a higher signal to noise ratio (SNR), as shown in Fig\ref{fig1}d (the bottom spectrum). The external magnetic vector field B can be monitored by observing the transition frequencies in lock-in signals. 
	
	\begin{widetext}
		\begin{minipage}{\linewidth}
			\begin{figure}[H]
				\centering
				\includegraphics[width=0.96\linewidth]{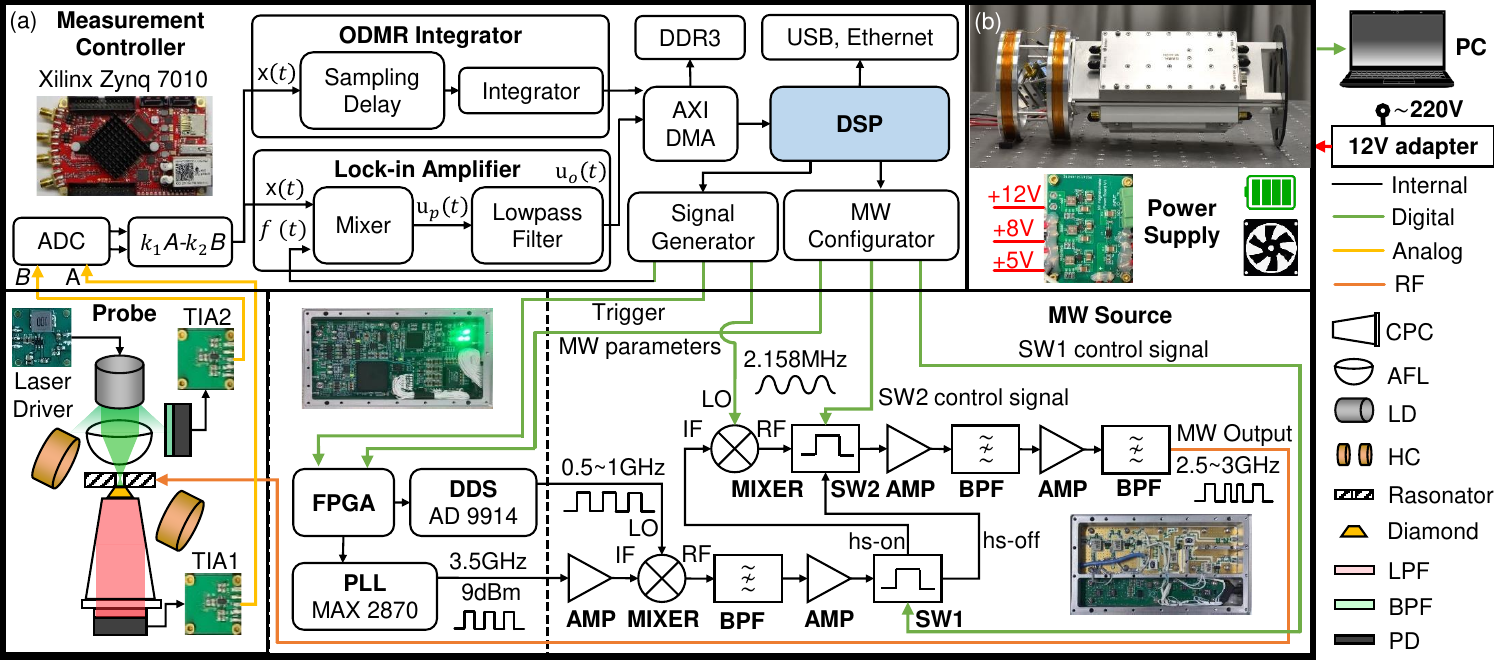}
				\caption{(a) Schematic of the presented magnetometer enclosed within the bold black frame, designed for connection to external devices such as a PC and a 12V adapter. The measurement controller transmits MW parameters and triggers the MW source to generate an adjustable base-band wave spanning 500-1000 MHz and a 3.5 GHz carrier wave. These two waves are mixed and amplified to produce a 2.5-3 GHz output. The MW source's "hs-on"/"hs-off" switch states control its operational modes, enabling the output of a multi-frequency microwave. The microwave is then conveyed by a resonator for spin manipulation. The fluorescence signal A from TIA1 and the scattered green light signal B from TIA2 are digitized by the ADC followed by balance detection. Subsequently, the signal undergoes an ODMR Integrator or a lock-in amplifier for data processing. (b) Diagram illustrating the overall structure of the magnetometer.}
				
				\label{fig2}
			\end{figure}    
		\end{minipage}
		
	\end{widetext}
	
	To realize the DNV magnetometer, we designed overall customized optics, on-board electronics and mechanical structures as shown in Fig. \ref{fig2}. Some components of them are based on our former works \cite{ref3.1.4,ref3.1.3,ref3.2.1}. The magnetometer is composed of a measurement controller, a microwave (MW) source and an integrated optical probe. The measurement controller based on a commercial field-programmable gate array (FPGA) board implements a lock-in amplifier and a embedded software system for signal management, data processing and communications, which are crucial to meet the specialized requirements of spin-ensemble-based quantum systems. The home-made MW source is realized by a direct digital synthesizer (DDS), whose output frequency (up to 1.4 GHz) is up-converted to the working range ($\sim$ 3 GHz), capable of fast frequency modulation with a bandwidth up to MHz-level. During the lock-in detection, the output of the MW source is modulated in a frequency-shift keying (FSK) mode, in synchronized with a digital signal generated from the measurement controller, and the shift of resonant frequency can be extracted from demodulation of the collected fluoresce from the probe. The integrated optical probe includes a high-power laser module and a high-efficiency fluorescence collection segment. The supporting components in the system are all home made or highly customized such as the power supply, photoelectric detector and laser driver.
	
	\begin{figure}[ht]
		\centering
		\includegraphics[width=0.48\textwidth]{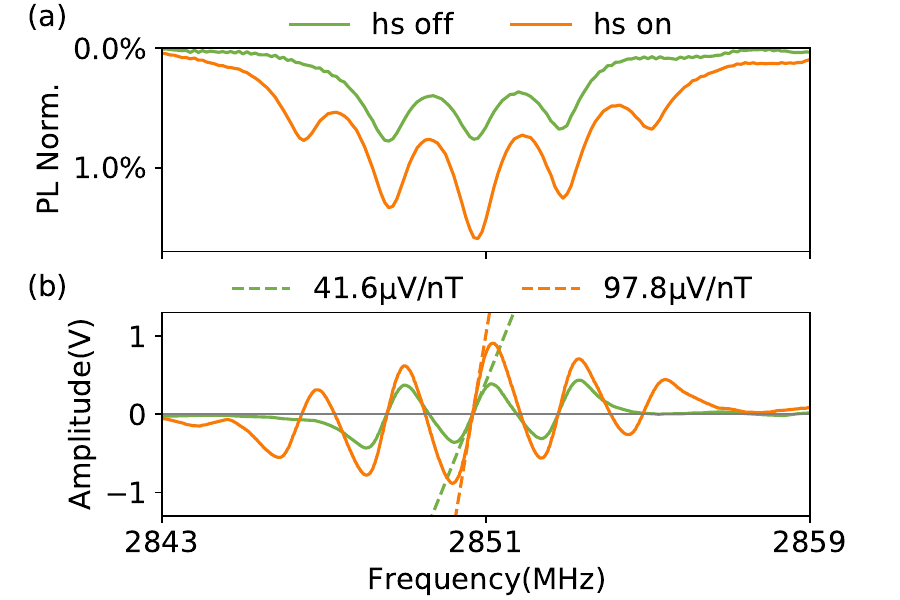}
		\caption{Simultaneous driving of hyperfine splitting resonances in (a) ODMR spectrum and (b) Lock-in demodulated signals. The notation "hs on(off)" duplicates the experiment use (not use) simultaneous driving of the spins. All experiments are performed using modulation frequency at 1 kHz and modulation depth at 400 kHz. We performed linear fitting to determine the steepest slopes around the zero point of each curve.}
		\label{fig3}
	\end{figure}

\begin{figure}[ht]
		\includegraphics[width=0.48\textwidth]{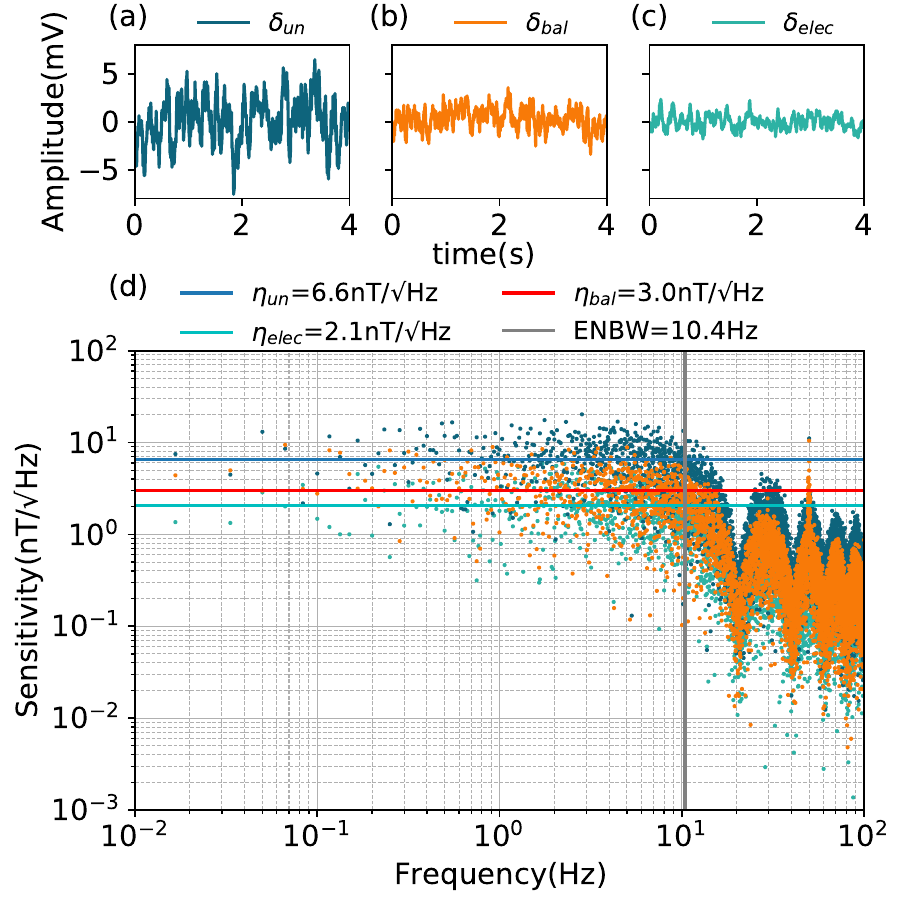}
		\caption{Sensitivity estimation of the magnetometer with amplification coefficients k$_1$ and k$_2$ set to 1000 and 930, respectively, to minimize noise. (a) The original noise at an off-resonant frequency of 2.54 GHz, including $\delta_{un}$  without balance detection, $\delta_{bal}$ with balance detection, and $\delta_{elec}$ with the laser and MW turned off. Each dataset was collected over a 60-second period. (b) Sensitivity values for the four measurements. The electronic noise sensitivity $\eta_{elec}$, balanced detection sensitivity $\eta_{bal}$, and unbalanced detection sensitivity $\eta_{un}$ are 2.1, 3.0, and 6.6 nT/$\sqrt{Hz}$, respectively, with an ENBW of 10.4 Hz.}
		\label{fig4}
	\end{figure}

	The magnetometer was tested in an unshielded environment, applying a commercial regulated power supply(Rigol DP832) and the sensitivities were all estimated at the smallest bias-field projected orientation using the MSD method \cite{refS1}, subject to the power supply's performance. As shown in Fig.\ref{fig1}d, the electron-spin resonance is split into three individual lines because of $^{14}$N hyperfine interactions. To excite all the spin signals, we introduce a 2.158MHz modulation on the final output microwave by a frequency mixer before radiated to the sample \cite{ref1.11.1,ref1.11.3,ref1.11.4,ref1.11.6}. By adjusting carefully the DC offsets of the mixer inputs, the output two mixed side-band together with the local oscillation (LO) forms a equally-spanned triple-frequency microwave. By simultaneously driving, we achieve a 2.3-fold enhancement in SNR as shown in Fig.\ref{fig3}, close to the theoretical expectation of 3. The sensitivity under this circumstance is around 5.5 nT/$\sqrt{Hz}$ as shown in Fig.\ref{fig4}d ($\eta_{un}$) with an an equivalent noise bandwidth (ENBW) of 10.4 Hz, while the photon-shot-noise-limited sensitivity is about 8.3 pT/$\sqrt{Hz}$, inferred from the half width at half maximum (HWHM, 617 kHz) and the signal contrast (1.53\%) of the ODMR spectrum. 
	
	To further optimize the performance, the balanced or differential photodetection scheme is often adopted, where nearly one-order of improvement has been reported before \cite{ref1.7.8,ref1.7.9,ref1.11.1,ref1.11.3}. However in our system, conventional balanced photodetection is not feasible due to the limited space. Instead, we utilize a digital scheme that detected the amplitudes of the fluoresce and background light independently as shown in Fig.\ref{fig2}, and then numerically compute the differential value associated with a amplification coefficient $k_i$ for each input. By adjusting the coefficients, a 2.2 times of SNR improvement is observed in Fig.\ref{fig4}a-b. After this optimization, we determined that with the same ENBW as above the sensitivity $\eta_{bal}$ reached 3 nT/$\sqrt{Hz}$ as shown in Fig.\ref{fig4}d.
	
	\begin{figure}[ht]
		\centering
		\includegraphics[width=0.48\textwidth]{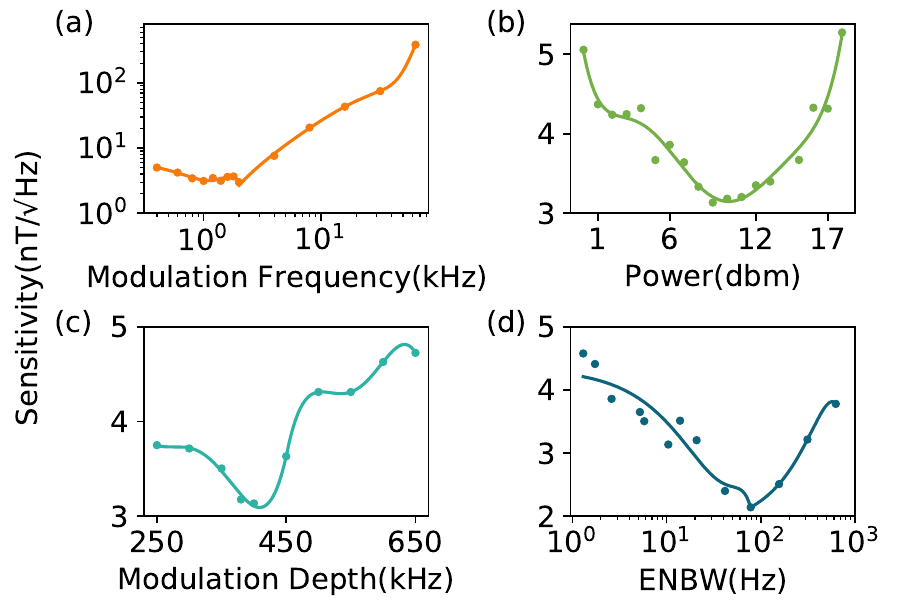}
		\caption{Performance under different experimental parameters, with a sampling rate of 200 Hz. (a)-(c) Variation of the modulation frequency from 0.4 kHz to 64 kHz, the microwave power from 0 dBm to 18 dBm and the modulation depth from 250 kHz to 650 kHz with the ENBW of 10.4Hz. (d) Variation of the ENBW from 1.3 Hz to 625 Hz, encompassing its scope, which is equivalent to an integration time from 0.9 ms to 384 ms.}
		\label{fig5}
	\end{figure}
	
	Finally, we assessed the magnetometer's performance under various experimental parameters, as depicted in Fig.\ref{fig5}a-d. We systematically adjusted one parameter at a time while keeping the others constant. The results revealed that the modulation frequency exerted the most significant influence on sensitivity. Reducing the modulation frequency from 32kHz to 1kHz led to a remarkable more than 20-fold improvement. In contrast, adjusting power and modulation depth resulted in less than a two-fold improvement in sensitivity. We observed different sensitivities, spanning from 2.14 nT/$\sqrt{Hz}$ to 4.58 nT/$\sqrt{Hz}$, while modifying the ENBW from 1.3 Hz to 625 Hz. An appropriate ENBW can be selected for specific applications in our system.

	\section{Outlook}
	
	Comparing with previous DNV integration works those were mainly based on commercial electronics, we developed all the components from bottom elements and integrate them together to ensure a low SWaP. Clearly, such home-made electronics are relatively less sufficient than commercial equipment and limit the performance of the magnetometry, resulting in sensitivities in the range of several nT/$\sqrt{Hz}$, which falls significantly short of the photon-shot-noise limit of 8.3 pT/$\sqrt{Hz}$. As shown in Fig. \ref{fig4}d, the electronic noise-floor $\delta_{ele}=$ 2.1 nT/$\sqrt{Hz}$ is comparable with real sensitivity $\delta_{bal}=$ 3 nT/$\sqrt{Hz}$, which indicates the main contribution to the sensitivity is the electronic noise. This is because of two reasons, first the bit width of the ADC is 14, less than commercial LIAs usually used. Second, the home-made LIA uses fixed-point numbers for computation due to the lack of on-chip resources, resulting a loss of accuracy in the digital signal processing. As a comparison, a 18-bit high-performance commercial LIA with full-size float-point processing (Stanford Research Systems SR850) can help DNV magnetometer achieving a sensitivity as low as 15 pT/$\sqrt{Hz}$\cite{ref1.1.9}.
	
	There're three extra problems in the system that can be improved in the future. First the the biggest improvement for the magnetometer could be obtained by improving the homogeneity of the microwave field. Many work proved that using three-dimensional resonators \cite{ref5.1.1,ref5.1.2,ref5.1.3,ref5.1.4,ref5.1.5} can provide a larger and more homogenous magnetic field than planar resonators, for example, a self-resonant microhelix can improve SNR by 28 times \cite{ref5.1.1}, and optical antennas, while radiating MW, can double the collected photon rate \cite{ref5.1.2}. Second the digital balance detection utilized scattered green light as a reference and could not eliminate the noise generated by the photodetector itself. Therefore, it only increased the sensitivity by a factor of 2.2. However, using a beam splitter and performing hardware balance detection can increase the sensitivity to more than 10 times \cite{ref1.11.1, ref1.11.3, ref1.7.9, ref1.7.8}. Third the substantial heat generated by the laser diode significantly affects the output light's stability. We have tested the probe using a commercial laser module with the same power, where a 27\% enhancement of performance is observed. Based upon the above improvements in the future, the expected sensitivity of the present DNV magnetometer would be on the order of hundreds of pT/$\sqrt{Hz}$, with a ENBW of 10$\sim$100 Hz specifically.

	
	\section{Discussion and Conclusion}
	Additional avenues to improve the performance or extend the functionality of DNV magnetometer can also be incorporated to our system. Applying flux concentrators is a viable option, particularly as space resolution is not a critical factor in our application scenarios. Significant improvements of ×254 \cite{ref1.7.3} and ×472 \cite{ref1.7.1} have been demonstrated. Another worthwhile consideration is the adoption of a light-trapping geometry, aiming to achieve a 5$\%$ conversion efficiency of pump photons into ODMR \cite{ref1.7.17}. Also, simultaneous vector magnetometry can also be introduced to measure four projective magnetic fields concurrently \cite{ref1.7.9}. Lastly, considering pulsed sensing protocols, which generally offer higher sensitivity compared to continuous-wave methods, presents another alternative for improvement \cite{ref4,ref1.1.9}.

	In summary, we have introduced a fully integrated DNV magnetometer with a sensitivity of up to 2.14 nT/$\sqrt{Hz}$. This magnetometer includes all the essential functionalities of a conventional platform, and the performance of key components is on par with commercial devices. We overcame the limitations of previous fully-integrated DNV magnetometers by designing and integrating an on-board lock-in amplifier, a DDS-based fast MW source, and an optical probe with high power laser. To optimize our system's performance, we simultaneously addressed three hyperfine splitting features through multi-frequency microwaves and implemented digital balance detection. We also adjusted the experimental parameters to estimate the magnetometer's performance. Finally, we discussed our limitations and feasible improvements. This DNV magnetometer, with a diameter of 13 cm and a length of 26 cm, is purposefully designed to be compatible with small UAVs, making it practical for a wide range of applications, such as navigation and surveying.
	
	\section{METHOD}
	\label{method}
	
	The measurement controller \cite{ref3.1.4,ref3.1.3} is based on a Xilinx Zynq 7010 SoC, featuring a dual-core processor (ARM Cortex-A9), a programmable logic unit (Xilinx Artix-7 FPGA), and on-chip memories. The SoC provides control over several high-speed peripherals, including a 14-bit dual-channel ADC (LTC2145CUP-14) operating at a maximum speed of 125 MSPS, DDR3 memory, direct memory access (DMA) with a standard advanced extensible interface (AXI), and standard communication interfaces like USB and Ethernet. These utilities enable the magnetometer to perform functions such as digital balance detection, lock-in amplification, and communication with external terminals. The MW source \cite{ref3.2.1} is developed based on a DDS chip (ADI AD9914), supporting frequency modulations with a bandwidth of up to 1.4 GHz. It incorporates radio frequency (RF) chips instead of individual components to reduce the system's overall volume. These components include PLLs (Maxim MAX2870), frequency mixers (Mini-Circuits MCA1-60LH+), power amplifiers (Mini-Circuits GVA-63), band-pass filters (BPF, BFCN-2900+), and high-speed switches (Mini-Circuits ZASWA-2-50DRA+).
	
	The green laser is generated by a high-power laser diode (LD, NICHA NUGM04) with a home-made driving circuit (TI TPS54200) at an output power around 450 mW. About half of the power (225 mW) passes through an aspheric condenser lens (AFL, IBTEK AC4301-A) and one third of the total power (175 mW) is focused on the diamond sample mounted on a double split ring resonator for microwave radiation \cite{ref5.1.6}. The sample (Element Six, 1.5×1.5×0.5 mm, 100 face) with 4.5-ppm NV concentration is specially cut at a 45° angle and polished on each side. It is positioned at the center of a Helmholtz coil (HC) to facilitate the application of a bias magnetic field. The fluorescence emitted from the diamond goes through a compound parabolic concentrator (CPC, Edmund $\#$17-709) and a long-pass filter (LPF, Edmund Stock $\#$15-448), and is collected by a photodetector. The photodetector consists of a photodiode (PD, HAMAMATSU s13955-01) with a photon sensitivity of 0.52A/w at 700nm wavelength and a trans-impedance amplifier (TIA, ADI AD8065) with an amplification of 200. Another photodetector with an amplification of 2200 is positioned to capture the scattered green light and these two output are provided to the ADC for the digital balance detection.
	
	\begin{acknowledgments}
		This work was supported by the Fundamental Research Funds for the Central Universities (Grant No. 226-2023-00139, 226-2023-00137), the National Natural Science Foundation of China (Grant Nos. 92265114, 92265204) and the Major Scientific Project of Zhejiang Laboratory (Grants No. K2023MB0AC09).
	\end{acknowledgments}

	\renewcommand{\refname}{REFERENCES}
	\nocite{*}

\end{document}